\documentclass[doublecol]{epl2} 
\usepackage{graphicx}
\usepackage[skip=2pt,font=normalsize]{caption}
\usepackage[backend=biber, style = phys, backref=true
]{biblatex}
\usepackage{amsmath}%,amssymb}
\usepackage{mathabx}
\usepackage{commath}
\usepackage{dsfont}
\usepackage{microtype}
\usepackage{flexisym}
\usepackage{hyperref}
\usepackage[export]{adjustbox}
\newcommand{\Dr}{D_\text{r}}
\newcommand{\Dt}{D_\text{t}}

\graphicspath{{submit_figures/}}
\title{Force-Free and Autonomous Active Brownian Ratchets}
\shorttitle{Autonomous Force-Free Active Brownian Ratchets}

\author{Constantin Rein \inst{1} \and Martin Kol{\' a}{\v r} \inst{2} \and Klaus Kroy \inst{1} \and Viktor Holubec \inst{2}}
\shortauthor{C. Rein \etal}

\institute{                    
  \inst{1} Leipzig University, Faculty of Physics and Earth Sciences, Institute for Theoretical Physics, Brüderstraße 16, 04081 Leipzig \email{rein@itp.uni-leipzig.de} \email{klaus.kroy@uni-leipzig.de}\\
  \inst{2} Charles University, Faculty of Mathematics and Physics, Department of Macromolecular Physics, V Holešovičkách 2, CZ-180 00 Praha \email{viktor.holubec@mff.cuni.cz }\\
}

\abstract{
Autonomous active Brownian ratchets  rectify active Brownian particle motion solely by means of a spatially modulated but stationary activity, without external forces. We argue that such ratcheting requires at least a two-dimensional geometry. The underlying principle is similar to the ratcheting induced by steric obstacles in microswimmer baths:  suitably polarized swimmers get channeled, while the others get trapped in low-activity regions until they loose  direction. The maximum current is generally reached in the limit of large propulsion speeds, in which the rectification efficiency  vanishes. Maximum efficiency is attained at intermediate activities and numerically found to be on the order of a few percent, for ratchets with simple wedge-shaped low-activity regions.
}

\begin{document}

\maketitle

\section{Introduction}
Brownian ratchets are subtle microscale transport devices~\cite{Peskin1993,Reed2003}. They combine two effects that individually do not promote  directed transport, namely isotropic Brownian motion and asymmetric environments, such that a net directed particle current is produced~\cite{Astumian1994,reimann_brownian_2002,Parrondo2002}. Conventional designs with passive particles  usually break the spatial symmetry by imposing an asymmetric potential. They also introduce a non-equilibrium element, namely an isotropic and often periodic driving mechanism that, by itself, does not introduce any directionality~\cite{reimann_brownian_2002}. Typical examples comprise the rocking (or ``flashing'') of the potential or the overall temperature~\cite{magnasco_forced_1993,Astumian1994,rousselet_directional_1994,reimann_brownian_2002}. More complex temperature fields have also been investigated~\cite{smoluchowski1912,FeynmanLectures,Buttiker1987,Landauer1988,Ryabov2016}. 

The self-propulsion of an active Brownian particle (ABP) represents yet another non-equilibrium mechanism that one ought to be able to exploit for ratcheting. While it does transiently break the spatial and temporal symmetry of equilibrium Brownian motion~\cite{falasco_2016}, it does not, by itself, give rise to a net macroscopic current. One would however expect that one of the simplest realizations of an active Brownian ratchet should consist of an ABP exposed to a spatially asymmetric (periodic) activity landscape. Yet, even though a number of ratchet designs with active particles have been discussed in the literature~\cite{angelani_2009,Leonardo2010,angelani_geometrically_2010,Angelani2011,ghosh_self-propelled_2013,Ai2017,Geiseler_2017_entropy,Geiseler_2017_nature,Geiseler_2017_PRE,Merlitz_2018,Pietzonka_2019}, none of them was based solely on a stationary activity landscape. Instead, some relied on ABPs placed in a soft potential in one spatial dimension~\cite{Angelani2011,Ai2017}, or in  asymmetric hard potentials in two-dimensions~\cite{angelani_2009,Leonardo2010,angelani_geometrically_2010,ghosh_self-propelled_2013, Pietzonka_2019}. The asymmetric potentials, so typical of conventional ratchets, can be relinquished entirely, though, if one exploits the tendency of ABPs to polarize towards low-activity regions and accumulate there~\cite{Sharma_2017,Merlitz_2018,Auschra2021,auschra_density_2021,soker_how_2021}. The standard flashing  potential can then be replaced by a dynamic activity landscape. Examples include propagating optical activation pulses that induce aligned or anti-aligned drifts, depending on the persistence length of the ABP motion relative to the pulse width and propagation speed \cite{Geiseler_2017_entropy,Geiseler_2017_nature,Geiseler_2017_PRE}. In general, traveling activity waves induce traveling density and orientation waves of the ABPs, and can thus plainly be employed to sort ABPs, e.g., by size \cite{Merlitz_2018}. 

To sum up, ratcheting has been demonstrated for active particles in spatially asymmetric potential landscapes or in space-and-time dependent activity landscapes. However, no fundamental symmetry prevents ABPs from ratcheting also in \emph{stationary} spatially asymmetric activity landscapes. In the following, we show that such ratchets are indeed realizable and explore the maximum current and rectification efficiency of a  class of simple shapes, numerically.

\section{Model}

We consider the motion of an ABP in a unit-square arena (thus taking its size as the natural length unit) with periodic boundary conditions in two dimensions (see Fig.~\ref{fig:sketch+polarization}). The state at time $t$ is fully characterized by the position ${\bf{r}}(t) = [{x}(t),{y}(t)]$ and polarization $\mathbf{n}(t)=[\cos\theta(t),\sin\theta(t)]$ of the ABP. The translational and rotational Brownian motion are represented by mutually independent and unbiased ($\left<\eta_i\right> = 0$) Gaussian white noises $\eta_i(t)$ of unit strength,  $\left<\eta_i(t)\eta_j(t')\right> = \delta_{ij}\delta(t-t')$, and diffusion constants $\Dt$ and $\Dr$, respectively. The stationary activity landscape enters via a superimposed deterministic speed field $v[x,y]$.  The dynamical equations for the ABP read
\begin{subequations}\label{eqn:general_LV_eqns}
\begin{align}
    \dot{x} &= v(x,y)\cos(\theta) + \sqrt{2\Dt} \eta_x,
    \label{eq:x}\\
    \dot{y} &= v(x,y) \sin(\theta) + \sqrt{2 \Dt} \eta_y,
    \label{eq:y}\\
    \dot{\theta} &= \sqrt{2 \Dr} \eta_\theta,
    \label{eq:Dr}
\end{align}
\end{subequations}
We only consider activity fields symmetric in the $y$-direction, $v(x,1/2+y)=v(x,1/2-y)$, so that $\left< \dot{y}(t)\right>=0$ and the steady-state current is a scalar $I=\left< \dot{x}(t)\right>$. Whenever $I$ is consistently positive or negative, the device exhibits ratcheting.  

A few general observations about the dynamics are gleaned directly from the above equations. First, the essential stochastic ingredient of the model is the rotational diffusion. If $\Dr$ is taken to infinity, the ABP motion looses its persistence.  The model then reduces to a passive gas locally equilibrated at a spatially modulated (effective) temperature $T=\Dt +v^2/2\Dr$, with Boltzmann's constant and the friction coefficient set to unity. While such a gas can move thermophoretically in the presence of a temperature gradient, it cannot maintain a steady current in a periodic temperature profile. (We comment on the more subtle limit of a Knudsen gas~\cite{Knudsen:1909,Steckelmacher:1986}, at the end of the paper.)  The ratcheting effect must thus entirely result from a clever combination of the more or less persistent motion in the high- and low-activity regions, respectively. 

For conceptual purposes, it is sufficient to consider discrete landscapes with $v$ being represented by a step function, since the dynamics is anyway low-pass filtered by the translational diffusion process. Any small scale details and discontinuities in $v(x,y)$ will thereby effectively be washed out. Also notice that setting the maximum value of $v$ to a very large (formally infinite) value  amounts to the idealization of strictly ballistic dynamics in the high-activity (or simply ``active'') regions.  Similarly, retaining a non-vanishing $\Dt>0$ to avoid an absorbing state, the minimum value of $v$ can safely be set to zero in the low-activity (or simply ``passive'') regions, without much loss of generality. This choice, which shall be adopted for the remainder, simply amounts to purely diffusive dynamics, inside the passive region. Notice that a  periodic  two-step function of a single scalar variable is necessarily symmetric. In one dimension, one thus clearly cannot achieve  autonomous ratcheting with a corresponding stationary activity field---nor (as shown below) with any other.  

In summary, translational diffusion acts as a regularization for step-wise constant activity profiles, so that the archetypal activity landscape discretely jumps between $v=0$ and some finite or possibly even infinite value $v$. In the latter case, the active region is traversed in no time, so that, the total dwell time $\tau$ of the particle in the unit cell is equal to the time spent in the passive region. The latter is independent of $v$ and, at first sight, of $\Dr$. However, $\Dr$ limits the ``take-off'' of ABPs emerging from the passive region, and in fact also the whole particle distribution at the active-passive boundary. For example, the ABP cannot take off if it emerges with a swim direction pointing back into the passive region. Also it can ``tunnel'' through narrow edges of the passive region.
One therefore generally still expects the current $I\simeq \tau^{-1}$ (in our unit length setup) and the dwell time $\tau$ to  depend on $\Dt$ and $\Dr$, even if one takes $v\to \infty$, in the active region.   It is however plausible, and indeed corroborated by our Brownian dynamics simulations of the model presented below that for a given geometric shape of the passive region, one can often find an optimum choice of $\Dt\simeq \Dr$ (Fig.~\ref{fig:dimless_problem_grand_plot}). Then $\tau(\Dt,\Dr) \to \tau(\Dr)$ depends solely on $\Dr$, implying $I\simeq \Dr$, with a purely geometric prefactor. The latter can only depend on dimensionless features of the shape (such as the parameters $\delta$ and $\varepsilon$ in Fig.~\ref{fig:sketch+polarization}). In other words, under such idealized conditions, the task of an optimum ratchet design is entirely reduced to a geometric optimization problem.

These general considerations based on an infinite step function $v(x,y)$ may not always be practically useful, from an active-matter perspective. For instance, an experimental realization of our idealized ABP might possibly only allow for a maximum speed $v$, below the asymptotic regime alluded to above (in which the dwell time in the arena equals the trapping time in the passive region). This will clearly reduce the ratchet current from its maximum value, and the dwell time will depend both on $\Dt\simeq \Dr$ and the maximum attainable value of $v$. This ``attenuated'' transport regime, with $\Dt\simeq \Dr\simeq v$ may be of particular practical interest, if the active speed of the ABP is regarded as a costly input. The most desirable \emph{modus operandi} of the ratchet will then not anymore be that of maximum current $I\simeq \Dr$, obtained in the limit  $v\to \infty$, because the ratio $I/v$ vanishes in this limit.  Instead, one will then typically be interested in conditions that optimize this ratio,  which can be interpreted as the rectification efficiency of the active ratchet, very much in the spirit of ABP engines and bacterial motors \cite{Leonardo2010,angelani_geometrically_2010,Pietzonka_2019}. The interested practitioner will then generally have to find the corresponding optimum parameter values $\Dt$, $\Dr$, and $v$ for a given ratchet geometry, numerically. 

The remainder of the paper is dedicated to a more comprehensive analysis of the above general considerations. In particular, we want to clarify  why stationary active Brownian ratchets can only be realized in at least two space dimensions. We also estimate realistic values of the maximum dimensionless current $I(v\to\infty)/\Dr$  and rectification efficiency $I/v$, for a simple wedge geometry. 

\section{One-dimensional activity patterns}

Already in one spatial dimension, spatially varying activity profiles
accommodate non-intuitive effects. For example, the mean first passage time may depend non-monotonically on the distance from a target and the target finding probability can increase if the activity increases towards the target~\cite{Vuijk_2018}. This seemingly contradicts the known fact that active particles spend less time in regions of higher activity. However, while the latter is a steady-state property, the former relates to transient behavior. In fact, when an ABP is oriented along an activity gradient, it accelerates and thus increases its chance to reach a target before it looses its orientation. Similarly, an ABP placed in the middle of a one-dimensional domain with a linear activity gradient reaches the high-activity end faster and more often than the low-activity end~\cite{Ghosh_2015}.  
Although these effects look promising with regard to designing autonomous active Brownian ratchets, e.g., with a saw--tooth-shaped stationary activity landscape, there is a catch.  In the cited experiments~\cite{Ghosh_2015,Vuijk_2018},  the state is repeatedly reset externally, by placing the particle back in its initial position upon reaching the target or the boundary of the arena. For a genuine ratchet, such \emph{``deus-ex-machina''} type outside interventions are clearly not a permissible option. 

More formally, one can demonstrate the absence of ratcheting in one-dimensional activity landscapes, as follows. Activity landscape can sort and locally accumulate ABPs according to their orientation, but they do not reorient them. Crucially, and quite in contrast to potential landscapes, activity landscapes do not exert any forces or torques on the ABPs, which are a crucial mechanism underlying the ratcheting of ABPs in one-dimensional potential landscapes~\cite{Angelani2011}. As all orientations are thus equally probable in an unbiased ensemble, the spatially integrated total polarization must vanish. Together with the continuity equation for particle number conservation~\cite{Herrmann_2020}, this entails that the net current vanishes, too. 
More concretely, one may evoke the continuity of the local polarization profile as a function of position, for piece-wise continuous activity profiles~\cite{soker_how_2021,auschra_density_2021,Auschra2021}. From this one concludes that, for a vanishing total polarization, there must be   at least be one position $x_0$ in the polarization profile at which the time-averaged orientation vanishes. The time averaged current $I$ at this point is given by the time-integral over $v[x(t)=x_0]\cos \theta(t)$. Up to a constant factor, this is just the vanishing time-averaged orientation. And since, in one spatial dimension, the continuity condition implies that the steady state current is spatially constant, $I$ vanishes  everywhere if it vanishes locally, at $x_0$. 
We have corroborated this conclusion by extensive Brownian dynamics simulations and by numerical solution of the Fokker--Planck equation, associated with Eq.~(\ref{eqn:general_LV_eqns}), using the method of Ref.~\cite{Holubec2019}.  

\section{Two-dimensional activity patterns}
\label{sec:2D_top_ratchet}

\begin{figure}
    \begin{center}
    \includegraphics[width=0.9\linewidth]{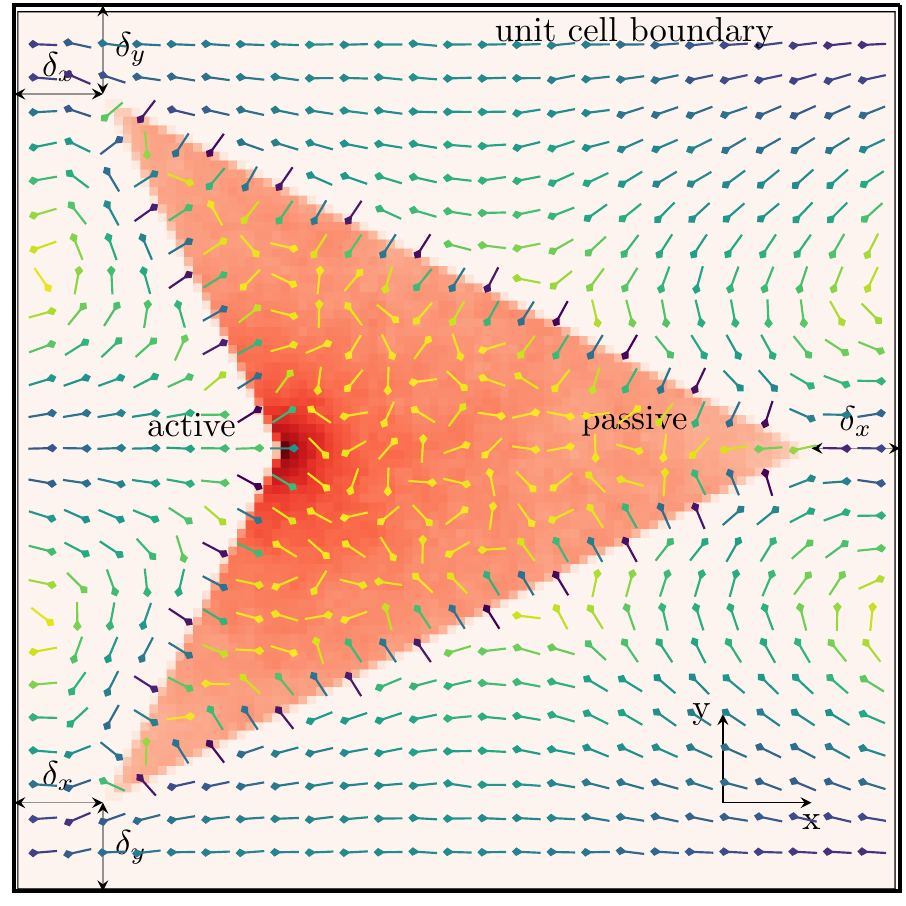}\\%
    \includegraphics{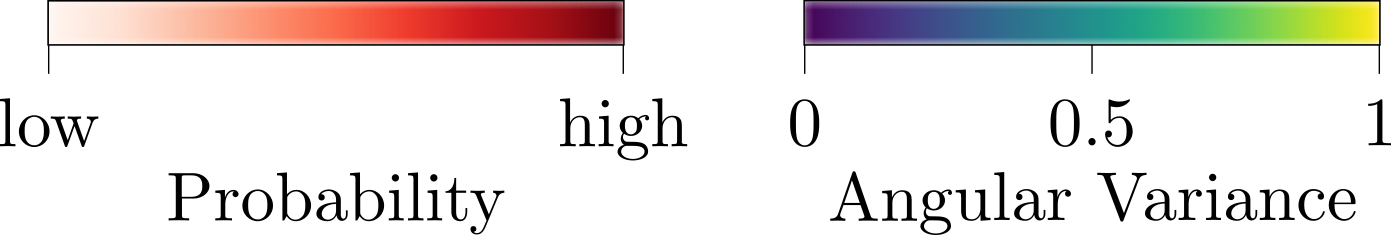}
    \end{center}
    \caption{Unit cell of a (unit width) two-dimensional square ratchet with $\delta_x=\delta_y=0.1$, $\varepsilon=0.75$, $v=\Dr$, and $\Dt=10^{-4}\Dr$. The background color encodes the probability density for the position of the ABP that predominantly dwells in the wedge-shaped passive region. Arrows show the mean orientation $\left<\mathbf{n}\right>$ of the ABP obtained from Brownian dynamics simulations, colors coding for the angular variance $1 - \Bigl(\left<n_x\right>^2 + \left<n_y\right>^2\Bigr)^{1/2}$; small values indicate strong alignment and {\cal O}(1)-values a random orientation.}
    \label{fig:sketch+polarization}
\end{figure}

Compared to one-dimensional activity landscapes, the situation is much different in two and higher-dimensional activity landscapes. The main reason is that the inevitable zeroes of the polarization do now no longer constrain the overall current to vanish, unless they cover a whole vertical line $(x_0,\{y\})$. The latter is by no means required by the condition on an overall vanishing polarization. Around an isolated point of vanishing current, the resulting systematic flow field (or, equivalently,  polarization field) takes the form of a vortex, as seen in Fig.~\ref{fig:sketch+polarization}.  
The sorting and accumulation of ABPs according to their orientation along the $x$-direction, which is already possible in one-dimensional activity landscapes~\cite{auschra_density_2021,Auschra2021,soker_how_2021}, and exploited in non-stationary active Brownian ratchets~\cite{Geiseler_2017_entropy,Geiseler_2017_nature,Geiseler_2017_PRE,Merlitz_2018}, is now modulated along the second spatial direction $y$. A particle moving along the $y$-direction therefore experiences an effectively  time-modulated activity pattern along the transport direction $x$, which has a similar rectifying effect as a dynamical one-dimensional activity profile.

The stationary but spatially periodically modulated activity-landscape $v(x,y)$ shown in Fig.~\ref{fig:sketch+polarization}  provides a proof-of-principle example and serves as an instructive illustration of a working ratchet. It features a piece-wise constant activity field with a wedge-shaped passive region, where $v(x,y) = 0$, in an otherwise moderately active unit square with constant $v(x,y) = \Dr$. The landscape is asymmetric along the $x$-direction and mirror-symmetric along the $y$-direction.
The dimensionless numbers $\delta_x$,  $\delta_y$, and $w = \varepsilon (1-2\delta_x)$,  with $\varepsilon\in[0,1]$,
denote the distances of the edges from the periodic boundaries and the width of the wedge along its mirror-symmetry axis, respectively. The extreme geometries corresponding to an infinitely thin passive region ($\varepsilon=0$) and a convex, triangular passive region ($\varepsilon=1$) both yield sub-optimal ratchets.

While even this simple wedge model is not exactly solvable, its performance can qualitatively be understood, using simple physical arguments. First, the above-mentioned saturation of the ratchet current for infinite speed $v \to \infty$ in the active region is simply due to the fact that the time spent by the ABP in the active region becomes negligible compared to the time $\tau$
spent diffusing in the passive region. This limit is thus amenable to event-driven simulations. Below, we go one step further and exploit it to construct a simplified geometric toy model that can provide semi-analytical estimates for the ratcheting current. Unfortunately, as already pointed out above, the conceptually convenient large-speed limit is somewhat academic. The practitioner will be interested in more affordable, finite values of $v$. Therefore, one should also consider the rectification efficiency $I/v$, which is the current produced by the ratchet relative to that of a perfectly polarized ABP.

To understand the pertinence of the limits of infinite or vanishing diffusivities $\Dr$, $\Dt$, recall that ratcheting is all about the geometric rectification of stochastic motion. In the limit $\Dr\to 0$ (perfect persistence), the initial polarization is however entirely conserved, while the limit $\Dr\to \infty$ (vanishing persistence) corresponds to thermophoresis within an effective temperature field. So both limits do not correspond to genuine active ratcheting.  Similarly, passive regions, with vanishing speed $v=0$, would all become absorbing for $\Dt\to 0$, while in active regions with a finite $v<\infty$, $\Dt\to\infty$ would wipe out the persistent active motion. Again, both limits are irrelevant for the discussion of active ratcheting. And even though one could set $\Dt = 0$ without creating an absorbing state if a non-vanishing speed $v>0$ was maintained in the passive (or less active) region, this choice would be unnatural, as it requires passive regions with vanishing (or even ``small'') $v$ to be administratively forbidden. On the other hand, allowing for some finite $\Dt\lesssim \Dr$ is not very consequential for the transport in the (more) active regions, where it merely partially degrades the persistence induced by the activity. This  exposes $\Dt$ as a parameter of minor physical relevance except for its regularizing role in the passive regions.  There are however two more reasons for including a non-vanishing $\Dt$, in the discussion. Firstly, it will actually matter for the comparison to practical physical realizations of an ABP ratchet. And secondly, it also serves to regularize some fine-grained details of the ratchet geometry, thereby putting a limit on an otherwise potentially limitless ornamentation of the ratchet design that would in practice have to be cut off by a physical particle radius. In contrast to the indispensable finite rotational diffusivity $\Dr$, the translational diffusivity $\Dt$ thus plays a rather technical role, as a model regularization parameter.

In conclusion, a pertinent discussion of a stationary ABP ratchet in two dimensions is best conducted for finite diffusivities $\Dr$ and $\Dt$.
While $\Dt^{-1}$ may at first suggest itself as the natural time unit of the ratchet (its dwell time), it turns out that its physical impact can, for a conceptual analysis, effectively be taken largely out of the game. The trick is to set it to an optimum value that maximizes the rectification efficiency $I/v$. Our numerical analysis (see Fig.~\ref{fig:dimless_problem_grand_plot}) confirms the expectation that this ``best'' value is unique and on the order of $\Dr$, for the simple geometry shown in Fig.~\ref{fig:sketch+polarization}. Its physical origin may be understood from the role played by $\Dt$ for controlling the ABP's escape time from the passive region. As already pointed out, above, if $\Dt\gg \Dr$, the ABP will not have lost its polarization when it leaves the passive region, and therefore typically swim right back into it, unless that region is narrow enough to be traversed with a substantial (``tunneling'') probability. Additionally, the dominance of translational diffusion for $\Dt\gg \Dr$ will unduly degrade the persistence in the active region beyond the inevitable minimum, set by $\Dr$. In contrast, if $\Dt\ll \Dr$, the regularizing effect of the translational diffusion onto the absorbing state may become less than optimal, as the initial particle polarization will then have been lost long before the ABP reemerges from the passive region. Altogether, this suggests an optimum value of $\Dt$ on the order of $\Dr$, as indeed numerically confirmed in Fig.~\ref{fig:dimless_problem_grand_plot}.  

To summarize, the  natural length unit of the stationary active ratchet is set by the domain size, its natural time unit by the inverse rotational diffusion coefficient $\Dr^{-1}$. And it is conceptually convenient (if not generally highly advisable) to work with an optimized translational diffusivity $\Dt \simeq \Dr$ of comparable magnitude.  The natural scale for the maximum ratchet current $I \simeq \tau^{-1} \simeq \Dr$ is then $\Dr$ itself, while that of the natural efficiency $I/v$ is  $(\tau v)^{-1} \simeq \Dr/v$. In practice, both quantities may be expected to be somewhat reduced by a dimensionless geometrical shape factor. The crucial message is then that determining the optimum current $I/\Dr$ and efficiency $I/v$ boils down to an infinite dimensional geometric optimization problem intertwined with the ``thermodynamic'' optimizations of the parameters $\Dt$ and $\Dt$, $v/\Dr$, respectively. 

\section{Numerical study}
\label{sec:numerical_simulations}

To provide a specific but instructive example, Fig.~\ref{fig:sketch+polarization} illustrates the working principle of the active Brownian ratchet  and its polarization field $ \mathbf{n}(t)$ for a wedge-shaped passive region in the unit square, with periodic boundary conditions. As already alluded to above, the orientation field is indeed seen to form vortices around the points with vanishing average orientation, which help to defy the no-go theorem for one-dimensional active ratchets. To create the figure, we solved Eq.~\eqref{eqn:general_LV_eqns} by a Brownian dynamics simulation with time-step $dt=10^{-4}/v$. The central observable is the ratchet current $I = x(T)/T$, evaluated as the final traversed $x$-distance of the ABP divided by the total simulation time $T=10^{7}/v$.
We checked that the vertical current $y(T)/T$ in the $y$-direction vanishes, as expected.
As demonstrated in Refs.~\cite{soker_how_2021,Auschra2021,auschra_density_2021}, along the active--passive boundary, the ABP points on average towards the passive region. This may seem surprising, since it seems to imply a net particle influx into the passive region.  It is  an illusion, however, since the swim pressure acting onto  an active-passive boundary is not exerted across it~\cite{Auschra2021}.  Actually, the particle can therefore ``escape'' from the passive region, against this swim pressure. If it escapes along the tip-side (right in Fig.~\ref{fig:sketch+polarization}), it likely ends up in the indented concave part of the passive region (left in Fig.~\ref{fig:sketch+polarization}). On the other hand, if the ABP escapes in the vertical direction towards the horizontal active channels of width $2\delta_y$  (top and bottom in Fig.~\ref{fig:sketch+polarization}), it can generate a net current from right to left. As a result, the passive region blocks particle paths to the right more than those to the left. Remarkably, active Brownian ratchets relying on potential forces acting like hard walls~\cite{hulme_coli_2008,berdakin_quantifying_2013,ghosh_self-propelled_2013,Pietzonka_2019,angelani_2009,angelani_geometrically_2010} are based on the very same principle. The important difference here is that our setup does not involve any potential forces, and the ABP can thus freely pass back and forth between the passive and active region. With hard walls, the ABP would slide along the wedge until it gets trapped in the pocket or escapes into the channel, thereby generating a net ratchet current. In our force-free active ratchet, the sliding motion is replaced by the diffusive spreading inside the passive region.

\begin{figure}
\begin{center}
\includegraphics[width=\linewidth]{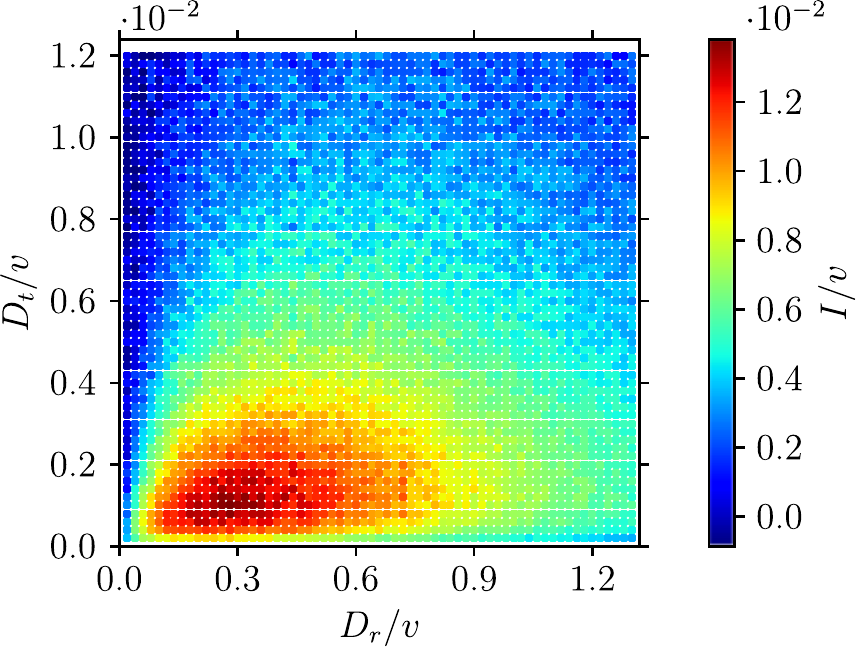}
\end{center}
    \caption{Rectification efficiency $I/v$ as function of the inverse P{\'e}clet numbers $\Dr/v$ and $\Dt/v$, for the active Brownian ratchet depicted in Fig.~\ref{fig:sketch+polarization}.}
    \label{fig:dimless_problem_grand_plot}
\end{figure}

For the setup illustrated in Fig.~\ref{fig:sketch+polarization}, we also investigated the rectification efficiency $I/v$ for finite activity, $v<\infty$, as a function of the diffusivities $\Dr$ and $\Dt$. In accord with our foregoing qualitative considerations, the numerical results shown in Fig.~\ref{fig:dimless_problem_grand_plot} feature a maximum around $I/v \sim 0.014$ for $\Dr \sim 0.3v$ and $\Dt \sim 0.001v$. These optimum values are specific for the chosen geometry and cannot be found without performing the numerical simulation. 

\begin{figure}
\begin{center}
    \includegraphics{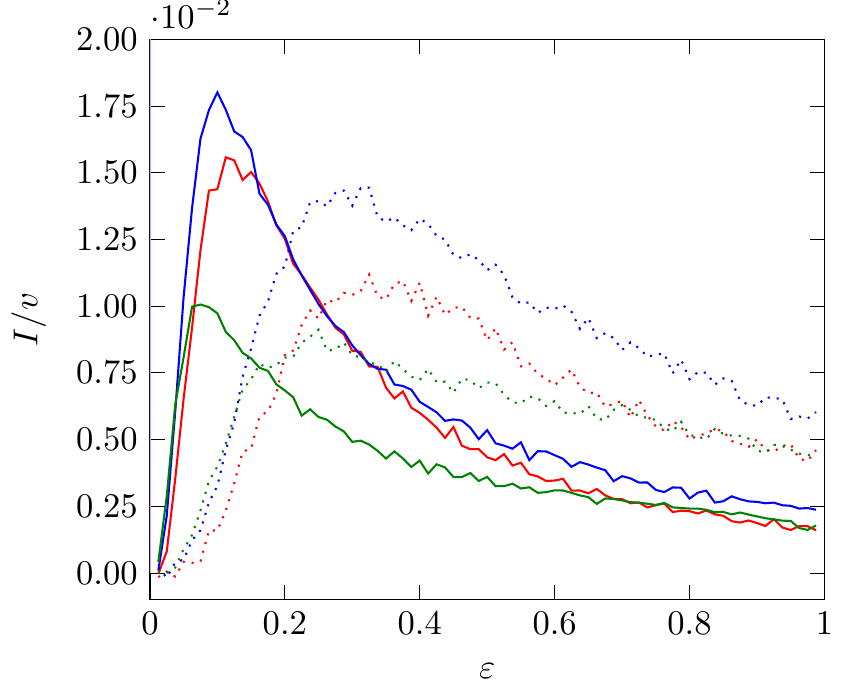}
\end{center}
 \caption{Rectification efficiency $I/v$ as a function of the indentation depth, parametrized by $1-\varepsilon$. Various combinations of $\Dr$ and $\Dt$ are shown, with colors coding for the value of $\Dr/v$: $0.1$ (red), $0.3$ (blue) and $1$ (green), and line style for $\Dt/v:$ $10^{-4}$ (solid) and $10^{-3}$ (dotted).}
    \label{fig:current_for_various_shapes}
\end{figure}

A more challenging task is to find the most efficient ratchet geometry. Here, we restrict this infinite dimensional optimization problem to the class of wedge or arrowhead shapes illustrated in Fig.~\ref{fig:sketch+polarization}. We ask for the optimum  depth of the concave indentation, which is parametrized by $\varepsilon$. For shallow indentations, the ABP spends more time in the passive region as needed to loose its polarization. This reduces the current and the rectification efficiency compared to a design with a stronger indentation. However, for very deep indentations, the passive region becomes too narrow to allow for a substantial reorientation of the traversing ABP, and the corresponding ``tunneling'' of the polarization eventually nullifies the ratcheting effect ($I \propto \varepsilon \to 0$). In other words, there is necessarily a non-monotonic dependence of the rectification efficiency on $\varepsilon$.  As illustrated in Fig.~\ref{fig:current_for_various_shapes}, this implies that the intermediate optimum value of $\varepsilon$, once again, needs to be found numerically. This result also nicely demonstrates the difference between our force-free active ratchet and its siblings operating with potential forces. In particular, for ratchets with hard walls around an exclusion zone of the same shape as our passive region, the ratcheting  would always be maintained, regardless of the wall thickness. The figure demonstrates that the non-monotonic dependence of the rectification strength on the indentation depth is robust against the fine tuning of the diffusivities, and that the optimization depends on the interplay between the geometry and the  inverse P{\'e}clet numbers $\Dt/v$ and $\Dr/v$. 
 
Beyond the indentation depth, one can also consider the effect of the parameter $\delta_y$ for the lateral width of the horizontal active channels. The current decreases both as $\delta_y \to 0$, when the channel width vanishes, and for $\delta_y \gtrsim 1/2$, when the passive volume becomes marginal relative to the overall domain size. Similarly, as for $\varepsilon$, $\Dr/v$ and $\Dt/v$, the rectification efficiency $I/v$ thus also exhibits a maximum as a function of $\delta_y$. Finally, the remaining parameter $\delta_x$ measures the overall width of the passive region in the $x$-direction. When $\delta_x\to 1/2$, the width of the passive region vanishes, and therefore also the current $I$, similarly as for $\varepsilon\to 0$. On the other hand, the current monotonically increases with decreasing $\delta_x\to 0$, until the passive region spans across the whole domain. 
Together, the shape parameters $\delta_x$, $\delta_y$, and $\varepsilon$ control how pointed and asymmetric the passive region may become. Generally speaking, $I/v$ grows with increasing asymmetry.

\section{Geometric toy model} \label{sec:geometric_toy_model}
\begin{figure}
    \centering
    \includegraphics[width=0.26\linewidth,valign=t]{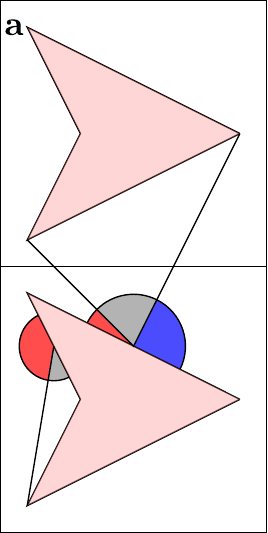}%
    \includegraphics[width=0.7\linewidth,valign=t]{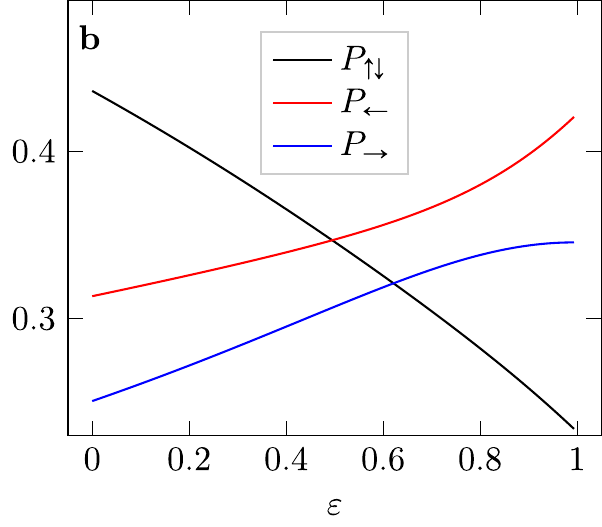}%
    \caption{Geometric toy model for the ratchet of Fig.~\ref{fig:sketch+polarization}; \textbf{a:} dependent on the emission site and direction from the passive region, the ABP contributes a current to the left (red), to the right (blue), or a ``neutral'' vertical current; \textbf{b:} overall probabilities for the traversals depicted in \textbf{a}.}
	\label{fig:prob_var_Bx}
\end{figure}

A more mechanistic insight into the effects of the ratchet geometry on the current can be obtained from a schematic, purely geometrical toy model. It is defined by the idealized rules that the particle moves with
infinite speed $v\to \infty$ in the active region and rotates and spreads 
sufficiently fast throughout the passive region to emerge from its surface with uniform spatial and orientational distributions, after a dwell time $\tau$. 
The ensuing simplifications enable us to bypass the computationally expensive Brownian dynamics simulations for qualitative estimates.
%\begin{enumerate}
%	\item \label{item_AR_vel} In the active region, $v_0 \rightarrow \infty$.
%	\item \label{item_unifrombdryprob_exit} Once the ABP enters the passive region, it will leave it after a fixed time $\tau$ from a random point at its boundary.
%	\item \label{item_unifromangleprob_exit} The orientation $\mathbf n$ of the ABP leaving the passive region is uniformly distributed within the allowed outward-pointing orientations.
%\end{enumerate}
The path of the ABP in the active region is then uniquely determined by the ratchet geometry alone. Once the ABP leaves the passive region with randomized orientation and position, it immediately hits either another part of the same passive region or one of its periodoc images, as sketched in Fig.~\ref{fig:prob_var_Bx}. 

One can therefore evaluate the probabilities $P_{\leftarrow}$, $P_{\rightarrow}$, and $P_{\updownarrows}$ that the ABP leaving the passive region travels to the left, right, or merely vertically, respectively. If the dwell time $\tau$ is approximated by the average reorientation time $ \tau = \Dr^{-1}$ of the ABP, as would be the case for an optimum choice of $\Dt$, one estimates  the current as
\begin{equation}
I =\Dr (P_{\leftarrow} -  P_{\rightarrow}).
\end{equation}
The resulting probabilities are shown in Fig.~\ref{fig:prob_var_Bx} as functions of the dimensionless horizontal width $\varepsilon$ of the symmetry axis of the wedge-shaped passive region. One sees that  $P_{\leftarrow} >  P_{\rightarrow}$ for all values of $\varepsilon$, so that the model always predicts a  leftward current $I$ that is numerically roughly comparable to the optimum currents obtained from the Brownian dynamics simulations. It naturally overestimates the current for extreme values of $\varepsilon$, corresponding to concave and vanishing passive volumes, respectively.  The actual reorientation of the ABP is then much less efficient than assumed by the stylized model, so that the comparison further corroborates the primary role played by the optimized destruction of the particle polarization in the passive region, for the rectificaton efficiency of the  ratchet.  

An iterative evaluation of the toy model provides further insight into the role played by the active channels separating the passive image regions. Starting from an initially uniform position distribution, one can find the distribution of positions where an ABP ensemble leaving the active-passive boundary will become trapped on the boundary again. The resulting position distribution can be used as the initial condition for the next step, again assuming uniformly distributed  orientations, for simplicity. After many iterations of this procedure, the position distribution no longer changes and one can consider it as an approximate stationary position distribution of the ABP. The resulting stationary distribution is similar to the one obtained from the Brownian dynamics simulations, depicted in Fig.~\ref{fig:sketch+polarization}. It exhibits a maximum in the indentation pocket of the passive region and, for $\delta_x=\delta_y=0$, also at the reverse indentations connecting the passive region with its periodic images. These particle accumulations hint at the sensitivity of ratcheting to the precise geometry, as they would leak out into the horizontal active channels to constitute the ratchet current, for $\delta_y>0$. 

Let us finally come back to the similarities and differences between our toy model and gases in similar geometries.  Dense gases or fluids, in which frequent mutual particle collisions can be relied on for establishing  local equilibrium, should not exhibit ratcheting in spatially periodic setups, like ours.
But in so-called rarefied or Knudsen~\cite{Knudsen:1909} gases, without an efficient local equilibration mechanism,   particles move  ballistically in the space between boundaries, similarly as ABPs in the active region of our geometric toy model, so that the analysis of transport largely boils down to the problem of  boundary conditions. This is then a more subtle issue~\cite{Steckelmacher:1986,Qiao2022} that would deserve further study.

\section{Conclusion}

Spatially inhomogeneous activity profiles can be used to sort active Brownian particles according to their orientations~\cite{soker_how_2021,Auschra2021,auschra_density_2021}. In one spatial dimension, the requirements for the overall system's polarization to vanish, together with particle conservation, prevent ratcheting in time-constant spatially-periodic activity landscapes. In two and more dimensions, such active ratcheting is possible. We analyzed a proof-of-principle realization of a wedge shaped two-dimensional autonomous force-free active Brownian ratchet. It demonstrates that active ratcheting does not require a dynamic activity profile, nor help from potential forces or walls.

Our study can be generalized in several ways. For example, it seems worthwhile to find out whether the wedge-shaped ratchet design maximizes the current or can be surpassed by more optimized geometries. Another potentially interesting extension could be to ABPs with translational and/or orientational inertia~\cite{Lowen2020}. And, eventually, it would be nice if the ratcheting currents in rarefied gases, hinted at by our toy model, could be experimentally demonstrated.

\section{Acknowledgements}
\acknowledgments
We acknowledge financial support by the pre-doc award program at Leipzig University, as well as by the Czech Science Foundation (project No.~20-02955J), and Charles University (project PRIMUS/22/SCI/009).
\printbibliography

@article{Qiao2022,
journal = {Phys. A: Stat. Mech. Appl.},
volume = {596},
pages = {127105},
year = {2022},
issn = {0378-4371},
doi = {https://doi.org/10.1016/j.physa.2022.127105},
url = {https://www.sciencedirect.com/science/article/pii/S0378437122001388},
author = {Yu Qiao and Zhaoru Shang}
}

@article{Lowen2020,
author = {L{\"o}wen,Hartmut },
journal = {J. Chem. Phys.},
volume = {152},
number = {4},
pages = {040901},
year = {2020},
doi = {10.1063/1.5134455},

URL = { 
        https://doi.org/10.1063/1.5134455
    
},
eprint = { 
        https://doi.org/10.1063/1.5134455
    
}

}

@article{Holubec2019,
  author = {Holubec, Viktor and Kroy, Klaus and Steffenoni, Stefano},
  journal = {Phys. Rev. E},
  volume = {99},
  issue = {3},
  pages = {032117},
  numpages = {18},
  year = {2019},
  month = {3},
  publisher = {American Physical Society},
  doi = {10.1103/PhysRevE.99.032117},
  url = {https://link.aps.org/doi/10.1103/PhysRevE.99.032117}
}

@article{falasco_2016,
  author = {Falasco, Gianmaria and Pfaller, Richard and Bregulla, Andreas P. and Cichos, Frank and Kroy, Klaus},
  journal = {Phys. Rev. E},
  volume = {94},
  issue = {3},
  pages = {030602},
  numpages = {5},
  year = {2016},
  month = {9},
  publisher = {American Physical Society},
  doi = {10.1103/PhysRevE.94.030602},
  url = {https://link.aps.org/doi/10.1103/PhysRevE.94.030602}
}

@article{Steckelmacher:1986,
doi = {10.1088/0034-4885/49/10/001},
url = {https://dx.doi.org/10.1088/0034-4885/49/10/001},
year = {1986},
month = {10},
publisher = {},
volume = {49},
number = {10},
pages = {1083},
author = {W Steckelmacher},
journal = {Rep. Prog. Phys.},
}

@article{Knudsen:1909,
  author={Knudsen, Martin},
  journal={Annalen der Physik},
  volume={336},
  number={1},
  pages={205--229},
  year={1909},
  publisher={WILEY-VCH Verlag Leipzig}
}

@article{Herrmann_2020,
  author = {Hermann, Sophie and Schmidt, Matthias},
  journal = {Phys. Rev. Res.},
  volume = {2},
  issue = {2},
  pages = {022003},
  numpages = {6},
  year = {2020},
  month = {4},
  publisher = {American Physical Society},
  doi = {10.1103/PhysRevResearch.2.022003},
  url = {https://link.aps.org/doi/10.1103/PhysRevResearch.2.022003}
}

@article{Leonardo2010,
author = {R. Di Leonardo  and L. Angelani  and D. Dell’Arciprete  and G. Ruocco  and V. Iebba  and S. Schippa  and M. P. Conte  and F. Mecarini  and F. De Angelis  and E. Di Fabrizio },
journal = {PNAS},
volume = {107},
number = {21},
pages = {9541-9545},
year = {2010},
doi = {10.1073/pnas.0910426107},
URL = {https://www.pnas.org/doi/abs/10.1073/pnas.0910426107},
eprint = {https://www.pnas.org/doi/pdf/10.1073/pnas.0910426107}
}

@Article{Peskin1993,
author={Peskin, C. S. and Odell, G. M. and Oster, G. F.},
journal={Biophys. J.},
year={1993},
volume={65},
number={1},
pages={316–324},
doi={10.1016/S0006-3495(93)81035-X},
url={https://doi.org/10.1016/S0006-3495(93)81035-X}
}

@Article{Landauer1988,
author={Landauer, Rolf},
journal={J. Stat. Phys.},
year={1988},
month={10},
day={01},
volume={53},
number={1},
pages={233-248},
issn={1572-9613},
doi={10.1007/BF01011555},
url={https://doi.org/10.1007/BF01011555}
}

@Article{Buttiker1987,
author={B{\"u}ttiker, M.},
journal={Zeitschrift f{\"u}r Physik B Condensed Matter},
year={1987},
month={6},
day={01},
volume={68},
number={2},
pages={161-167},
issn={1431-584X},
doi={10.1007/BF01304221},
url={https://doi.org/10.1007/BF01304221}
}

@article{Ryabov2016,
doi = {10.1088/1742-5468/2016/09/093202},
url = {https://dx.doi.org/10.1088/1742-5468/2016/09/093202},
year = {2016},
month = {9},
publisher = {IOP Publishing and SISSA},
volume = {2016},
number = {9},
pages = {093202},
author = {A Ryabov and V Holubec and M H Yaghoubi and M Varga and M E Foulaadvand and P Chvosta},
journal = {J. Stat. Mech. Theory Exp.}
}

@Article{Reed2003,
author={Reed, Steven I.},
journal={Nat. Rev. Mol. Cell Biol},
year={2003},
month={11},
day={01},
volume={4},
number={11},
pages={855-864},
issn={1471-0080},
doi={10.1038/nrm1246},
url={https://doi.org/10.1038/nrm1246}
}

@article{Ghosh_2015,
	%doi = {10.1103/physreve.92.012114},
	url = {https://doi.org/10.1103/Fphysreve.92.012114},
	year = 2015,
	month = {7},
	publisher = {American Physical Society ({APS})},
	volume = {92},
	number = {1},
	author = {Pulak K. Ghosh and Yunyun Li and Fabio Marchesoni and Franco Nori},
	journal = {Phys. Rev. E}
}

@Article{Ai2017,
author ="Ai, Bao-Quan and Li, Feng-Guo",
year  ="2017",
volume  ="13",
issue  ="13",
pages  ="2536-2542",
publisher  ="The Royal Society of Chemistry",
doi  ="10.1039/C7SM00405B",
url  ="http://dx.doi.org/10.1039/C7SM00405B"}

@article{Angelani2011,
	doi = {10.1209/0295-5075/96/68002},
	url = {https://doi.org/10.1209/0295-5075/96/68002},
	year = 2011,
	month = {12},
	publisher = {{IOP} Publishing},
	volume = {96},
	number = {6},
	pages = {68002},
	author = {L. Angelani and A. Costanzo and R. Di Leonardo}
}

@article{Sharma_2017,
  author = {Sharma, A. and Brader, J. M.},
  journal = {Phys. Rev. E},
  volume = {96},
  issue = {3},
  pages = {032604},
  numpages = {5},
  year = {2017},
  month = {9},
  publisher = {American Physical Society},
  doi = {10.1103/PhysRevE.96.032604},
  url = {https://link.aps.org/doi/10.1103/PhysRevE.96.032604}
}

@article{Merlitz_2018,
author = {Merlitz,Holger  and Vuijk,Hidde D.  and Brader,Joseph  and Sharma,Abhinav  and Sommer,Jens-Uwe },
journal = {J. Chem. Phys.},
volume = {148},
number = {19},
pages = {194116},
year = {2018},
doi = {10.1063/1.5025760},
URL = {https://doi.org/10.1063/1.5025760},
eprint = {https://doi.org/10.1063/1.5025760}
}

@Article{Geiseler_2017_entropy,
AUTHOR = {Geiseler, Alexander and Hänggi, Peter and Marchesoni, Fabio},
JOURNAL = {Entropy},
VOLUME = {19},
YEAR = {2017},
NUMBER = {3},
ARTICLE-NUMBER = {97},
URL = {https://www.mdpi.com/1099-4300/19/3/97},
ISSN = {1099-4300},
ABSTRACT = {Contrary to microbial taxis, where a tactic response to external stimuli is controlled by complex chemical pathways acting like sensor-actuator loops, taxis of artificial microswimmers is a purely stochastic effect associated with a non-uniform activation of the particles’ self-propulsion. We study the tactic response of such swimmers in a spatio-temporally modulated activating medium by means of both numerical and analytical techniques. In the opposite limits of very fast and very slow rotational particle dynamics, we obtain analytic approximations that closely reproduce the numerical description. A swimmer drifts on average either parallel or anti-parallel to the propagation direction of the activating pulses, depending on their speed and width. The drift in line with the pulses is solely determined by the finite persistence length of the active Brownian motion performed by the swimmer, whereas the drift in the opposite direction results from the combination of the ballistic and diffusive properties of the swimmer’s dynamics.},
%DOI = {10.3390/e19030097}
}

@article{ghosh_self-propelled_2013,
	volume = {110},
	issn = {0031-9007, 1079-7114},
	shorttitle = {Self-{Propelled} {Janus} {Particles} in a {Ratchet}},
	url = {https://link.aps.org/doi/10.1103/PhysRevLett.110.268301},
	%doi = {10.1103/PhysRevLett.110.268301},
	language = {en},
	number = {26},
	urldate = {2022-07-26},
	journal = {Phys. Rev. Lett.},
	author = {Ghosh, Pulak K. and Misko, Vyacheslav R. and Marchesoni, Fabio and Nori, Franco},
	month = 6,
	year = {2013},
	pages = {268301},
	file = {Submitted Version:/home/constantin/Zotero/storage/ANLDD594/Ghosh et al. - 2013 - Self-Propelled Janus Particles in a Ratchet Numer.pdf:application/pdf},
}

@article{berdakin_quantifying_2013,
	volume = {11},
	issn = {2391-5471},
	url = {https://www.degruyter.com/document/doi/10.2478/s11534-013-0300-7/html?lang=de},
	doi = {10.2478/s11534-013-0300-7},
	abstract = {Suitable asymmetric microstructures can be used to control the direction of motion in microorganism populations. This rectification process makes it possible to accumulate swimmers in a region of space or to sort different swimmers. Here we study numerically how the separation process depends on the specific motility strategies of the microorganisms involved. Crucial properties such as the separation efficiency and the separation time for two bacterial strains are precisely defined and evaluated. In particular, the sorting of two bacterial populations inoculated in a box consisting of a series of chambers separated by columns of asymmetric obstacles is investigated. We show how the sorting efficiency is enhanced by these obstacles and conclude that this kind of sorting can be efficiently used even when the involved populations differ only in one aspect of their swimming strategy.},
	language = {en},
	number = {12},
	urldate = {2022-11-23},
	journal = {Open Physics},
	author = {Berdakin, Iván and Silhanek, Alejandro and Cortéz, Hernán Moyano and Marconi, Verónica and Condat, Carlos},
	month = 12,
	year = {2013},
	%note = {Publisher: De Gruyter Open Access},
	keywords = {motility, ratchet, swimmer sorting},
	pages = {1653--1661},
	file = {Full Text PDF:/home/constantin/Zotero/storage/RBJ7XQ3V/Berdakin et al. - 2013 - Quantifying the sorting efficiency of self-propell.pdf:application/pdf},
}

@article{hulme_coli_2008,
	volume = {8},
	doi = {10.1039/b809892a},
	journal = {Lab chip},
	author = {Hulme, S and DiLuzio, Willow and Shevkoplyas, Sergey and Turner, Linda and Mayer, Michael and Berg, Howard and Whitesides, George},
	month = 12,
	year = {2008},
	pages = {1888--95},
}

@article{Geiseler_2017_PRE,
  author = {Geiseler, Alexander and H\"anggi, Peter and Marchesoni, Fabio and Mulhern, Colm and Savel'ev, Sergey},
  journal = {Phys. Rev. E},
  volume = {94},
  issue = {1},
  pages = {012613},
  numpages = {10},
  year = {2016},
  month = {7},
  publisher = {American Physical Society},
  doi = {10.1103/PhysRevE.94.012613},
  url = {https://link.aps.org/doi/10.1103/PhysRevE.94.012613}
}

@article{Geiseler_2017_nature,
  author={Geiseler, Alexander and H{\"a}nggi, Peter and Marchesoni, Fabio},
  journal={Sci. Rep.},
  volume={7},
  number={1},
  pages={1--9},
  year={2017},
  publisher={Nature Publishing Group},
  url={https://doi.org/10.1038/srep41884}
}

@article{Vuijk_2018,
  author = {Vuijk, Hidde D. and Sharma, Abhinav and Mondal, Debasish and Sommer, Jens-Uwe and Merlitz, Holger},
  journal = {Phys. Rev. E},
  volume = {97},
  issue = {4},
  pages = {042612},
  numpages = {7},
  year = {2018},
  month = {4},
  publisher = {American Physical Society},
  doi = {10.1103/PhysRevE.97.042612},
  url = {https://link.aps.org/doi/10.1103/PhysRevE.97.042612}
}

@article{Pietzonka_2019,
  author = {Pietzonka, Patrick and Fodor, \'Etienne and Lohrmann, Christoph and Cates, Michael E. and Seifert, Udo},
  journal = {Phys. Rev. X},
  volume = {9},
  issue = {4},
  pages = {041032},
  numpages = {21},
  year = {2019},
  month = {11},
  publisher = {American Physical Society},
  doi = {10.1103/PhysRevX.9.041032},
  url = {https://link.aps.org/doi/10.1103/PhysRevX.9.041032}
}

@article{reimann_brownian_2002,
	volume = {361},
	issn = {03701573},
	shorttitle = {Brownian motors},
	url = {http://arxiv.org/abs/cond-mat/0010237},
	doi = {10.1016/S0370-1573(01)00081-3},
	abstract = {Transport phenomena in spatially periodic systems far from thermal equilibrium are considered. The main emphasize is put on directed transport in so-called Brownian motors (ratchets), i.e. a dissipative dynamics in the presence of thermal noise and some prototypical perturbation that drives the system out of equilibrium without introducing a priori an obvious bias into one or the other direction of motion. Symmetry conditions for the appearance (or not) of directed current, its inversion upon variation of certain parameters, and quantitative theoretical predictions for specific models are reviewed as well as a wide variety of experimental realizations and biological applications, especially the modeling of molecular motors. Extensions include quantum mechanical and collective effects, Hamiltonian ratchets, the influence of spatial disorder, and diffusive transport.},
	language = {en},
	number = {2-4},
	urldate = {2022-11-23},
	journal = {Phys. Rep.},
	author = {Reimann, Peter},
	month = 4,
	year = {2002},
	%note = {arXiv:cond-mat/0010237},
	keywords = {Condensed Matter - Statistical Mechanics},
	pages = {57--265},
	annote = {Comment: Revised version (Aug. 2001), accepted for publication in Physics Reports},
	file = {Reimann - 2002 - Brownian motors noisy transport far from equilibr.pdf:/home/constantin/Zotero/storage/GTYCFX54/Reimann - 2002 - Brownian motors noisy transport far from equilibr.pdf:application/pdf},
}

@article{angelani_geometrically_2010,
	volume = {12},
	issn = {1367-2630},
	url = {https://dx.doi.org/10.1088/1367-2630/12/11/113017},
	doi = {10.1088/1367-2630/12/11/113017},
	language = {en},
	number = {11},
	urldate = {2022-11-23},
	journal = {New J. Phys.},
	author = {Angelani, Luca and Leonardo, Roberto Di},
	month = 11,
	year = {2010},
	pages = {113017},
}

@article{auschra_density_2021,
	volume = {103},
	issn = {2470-0045, 2470-0053},
	url = {http://arxiv.org/abs/2101.04548},
	doi = {10.1103/PhysRevE.103.062604},
	abstract = {Suspensions of motile active particles with space dependent activity form characteristic polarization and density patterns. Recent single-particle studies for planar activity landscapes identified several quantities associated with emergent density-polarization patterns, including swim pressure, that are solely determined by bulk variables and thus constitute state functions. We show that for radially symmetric activity steps these variables depend on the curvature of the active-passive interface. In the specific case of total polarization and swim pressure, we generalize this result for arbitrary radially symmetric activity landscapes. Our exact as well as approximate analytical results agree with exact numerical calculations. We expect qualitatively the same results to hold for arbitrarily curved activity landscapes.},
	language = {en},
	number = {6},
	urldate = {2022-07-26},
	journal = {Phys. Rev. E},
	author = {Auschra, Sven and Holubec, Viktor},
	month = 6,
	year = {2021},
	%note = {arXiv:2101.04548 [cond-mat]},
	keywords = {Condensed Matter - Soft Condensed Matter},
	pages = {062604},
	file = {Auschra and Holubec - 2021 - Density and Polarization of Active Brownian Partic.pdf:/home/constantin/Zotero/storage/3UIQSQ2G/Auschra and Holubec - 2021 - Density and Polarization of Active Brownian Partic.pdf:application/pdf},
}

@article{soker_how_2021,
	volume = {126},
	url = {https://link.aps.org/doi/10.1103/PhysRevLett.126.228001},
	doi = {10.1103/PhysRevLett.126.228001},
	abstract = {Active-particle suspensions exhibit distinct polarization-density patterns in activity landscapes, even without anisotropic particle interactions. Such polarization without alignment forces is at work in motility-induced phase separation and betrays intrinsic microscopic activity to mesoscale observers. Using stable long-term confinement of a single thermophoretic microswimmer in a dedicated force-free particle trap, we examine the polarized interfacial layer at a motility step and confirm that it does not exert pressure onto the bulk. Our observations are quantitatively explained by an analytical theory that can also guide the analysis of more complex geometries and many-body effects.},
	number = {22},
	urldate = {2022-07-26},
	journal = {Phys. Rev. Lett.},
	author = {Söker, Nicola Andreas and Auschra, Sven and Holubec, Viktor and Kroy, Klaus and Cichos, Frank},
	month = 6,
	year = {2021},
    %note = {Publisher: American Physical Society},
	pages = {228001},
}

@article{Auschra2021,
  author = {Auschra, Sven and Holubec, Viktor and S\"oker, Nicola Andreas and Cichos, Frank and Kroy, Klaus},
  journal = {Phys. Rev. E},
  volume = {103},
  issue = {6},
  pages = {062601},
  numpages = {20},
  year = {2021},
  month = {6},
  publisher = {American Physical Society},
  doi = {10.1103/PhysRevE.103.062601},
  url = {https://link.aps.org/doi/10.1103/PhysRevE.103.062601}
}

@Article{Parrondo2002,
author={Parrondo, J. M. R.
and de Cisneros, B. J.},
journal={Appl. Phys. A},
year={2002},
month={8},
day={01},
volume={75},
number={2},
pages={179-191},
issn={1432-0630},
doi={10.1007/s003390201332},
url={https://doi.org/10.1007/s003390201332}
}

@article{smoluchowski1912,
author = {Smoluchowski, M. v.}, 
journal = {Physik. Zeitschr.},
volume = {13},
year = {1912},
number = {1069.},
}

@book{FeynmanLectures,
author = {Feynman, R. P.  and Leighton, R. B. and Sands, M.},
volume = {1},
chapter = {46},
publisher ={Addison Wesley, Reading MA},
year = {1963},
}

@article{magnasco_forced_1993,
	volume = {71},
	issn = {0031-9007},
	url = {https://link.aps.org/doi/10.1103/PhysRevLett.71.1477},
	doi = {10.1103/PhysRevLett.71.1477},
	language = {en},
	number = {10},
	urldate = {2022-12-13},
	journal = {Phys. Rev. Lett.},
	author = {Magnasco, Marcelo O.},
	month = 9,
	year = {1993},
	pages = {1477--1481},
}

@article{Astumian1994,
  author = {Astumian, R. Dean and Bier, Martin},
  journal = {Phys. Rev. Lett.},
  volume = {72},
  issue = {11},
  pages = {1766--1769},
  numpages = {0},
  year = {1994},
  month = {3},
  publisher = {American Physical Society},
  doi = {10.1103/PhysRevLett.72.1766},
  url = {https://link.aps.org/doi/10.1103/PhysRevLett.72.1766}
}

@article{rousselet_directional_1994,
	volume = {370},
	issn = {0028-0836, 1476-4687},
	url = {http://www.nature.com/articles/370446a0},
	doi = {10.1038/370446a0},
	number = {6489},
	urldate = {2022-12-22},
	journal = {Nature},
	author = {Rousselet, Juliette and Salome, Laurence and Ajdari, Armand and Prostt, Jacques},
	month = 8,
	year = {1994},
	pages = {446--447},
}

@article{angelani_2009,
  author = {Angelani, Luca and Di Leonardo, Roberto and Ruocco, Giancarlo},
  journal = {Phys. Rev. Lett.},
  volume = {102},
  issue = {4},
  pages = {048104},
  numpages = {4},
  year = {2009},
  month = {1},
  publisher = {American Physical Society},
  doi = {10.1103/PhysRevLett.102.048104},
  url = {https://link.aps.org/doi/10.1103/PhysRevLett.102.048104}
}
\end{document}